\documentclass[a4paper,12pt%
              ]{article}
\usepackage{array}
\usepackage{epsf}

\newcolumntype{R}{@{}>{\displaystyle {}}r<{{}}}
\newcolumntype{L}{@{}>{\displaystyle {}}l<{{}}}

\newcommand{\tr}{\mathop{\mathrm{tr}}}
\newcommand{\Tr}{\mathop{\mathrm{Tr}}}
\newcommand{\rmd}{\mathrm{d}}

\renewcommand{\today}{October 16, 1995}

\newcommand{\vffb}[1]{\mbox{$
    \begin{array}{@{}*{2}{c@{}}}
      \vcenter{\hbox{\vphantom{\raisebox{0pt}[1.1\height]{\epsffile{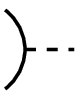}}}
               \epsffile{v3.eps}}}&#1
    \end{array}
$}}

\newcommand{\vffbb}[2]{\mbox{$
    \begin{array}{@{}*{2}{c@{}}}
      &#1\\
      \vcenter{\epsffile{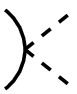}}\\
      &#2\\
    \end{array}
$}}

\begin{document}

\begin{titlepage}
  \renewcommand{\thefootnote}{\fnsymbol{footnote}}
  \begin{flushright}
    \begin{tabular}{l@{}}
      IASSNS-HEP-95/74\\
      HD-THEP-95-43\\
      HUB-EP-95/17\\
      hep-th/9510036
    \end{tabular}
  \end{flushright}

  \vskip 0pt plus 0.4fill

  \begin{center}
    \textbf{\LARGE Axial Couplings on the World-Line}
  \end{center}

  \vskip 0pt plus 0.2fill

  \begin{center}
    {\large
    Myriam Mondrag\'on%
    \footnote{E-mail: M.Mondragon@ThPhys.Uni-Heidelberg.de},
    Lukas Nellen%
    \footnote{E-mail: L.Nellen@ThPhys.Uni-Heidelberg.de},
    Michael G. Schmidt%
    \footnote{E-mail: M.G.Schmidt@ThPhys.Uni-Heidelberg.de}\\
    }
    \textit{
      Institut f\"ur Theoretische Physik,
      Philosophenweg 16,\\
      D-69120 Heidelberg,
      Germany
      }\\
    \vspace{1ex}
    {\large
    Christian Schubert%
    \footnote{E-mail: schubert@qft2.physik.hu-berlin.de}\\
    }
    \textit{
      Humboldt Universit\"at zu Berlin,
      Institut f\"ur Physik,\\
      Invalidenstr. 110,
      D-10115 Berlin,
      Germany}\\
    and\\
    \textit{
      School of Natural Sciences,
      Institute for Advanced Study,\\
      Olden Lane,
      Princeton, NJ 08540,
      USA
      }\\

    \vskip 1ex plus 0.3fill

    {\large \today}

    \vskip 1ex plus 0.7fill

    \textbf{Abstract}
  \end{center}
  \begin{quotation}
    We construct a world-line representation for the fermionic one-loop
    effective action  with axial and also vector, scalar, and pseudo-scalar
    couplings. We use this expression to compute a few selected scattering
    amplitudes. These allow us to  verify that our method yields the same
    results as standard field theory.
    In particular, we are able to reproduce the chiral anomaly.
    Our starting point is the second order formulation for
    the Dirac fermion. We translate the second order expressions into
    a world-line action.
  \end{quotation}

  \vskip 0pt plus 2fill

  \setcounter{footnote}{0}

\end{titlepage}

Recently it has been argued~%
\cite{str:npb385,ss:plb318,ss:plb331,ss:hd-thep-94-25,mck:ap224,mr:prd48}
that conventional
quantum field theory, at least in some problems, can be profitably
substituted by a first quantized formalism dealing with relativistic
particles on a world-line.  This is a very old
idea~\cite{fey:pr80,fey:pr84,hjs:prd16,pol:gfs},
but for a long time it was only applied to the
description of one-loop determinants and propagators, whereas in
ref.~\cite{ss:plb331, ss:hd-thep-94-25} a unified description for
whole classes of
higher loop Feynman graphs was proposed.  The Bern-Kosower
formalism~\cite{bk:prl66,bk:npb379,bd:npb379},
which triggered much of the recent research in this subject,
was derived from string theory.
It beautifully simplifies tree and one-loop calculations in field
theory but, unfortunately, becomes very difficult to develop for higher loops.
This is one of the reasons why Strassler's
observation~\cite{str:npb385} that their one-loop
rules can already be
derived in the first quantized relativistic particle approach, {\em
  i.~e.}, in a one-dimensional QFT on the world-line, is extremely
useful.  Indeed, already in the calculation of one-loop effective
actions this leads to impressive
simplifications~\cite{ss:plb318,fhss:hd-thep-94-26} compared to a
conventional heat kernel approach.  Higher
loop-calculations~\cite{ss:plb331,ss:hd-thep-94-25}, even though still
at a preliminary stage, look very promising.

In order to perform the analog of conventional QFT calculations it is
necessary to know the proper world-line Lagrangian.  Whereas for
scalar~$\phi ^4$ or~$\phi ^3$ potentials and vector gauge coupling to
scalar and Dirac particles this is well known, only very
recently we constructed a world-line Lagrangian~\cite{mnss:hd-thep-94-51} for
the
Yukawa couplings of scalars and pseudo-scalars to Dirac particles.

In this letter we present a general formalism for the simultaneous
coupling of abelian vector, axial vector, scalar, and pseudo-scalar
background fields
to Dirac particles in loops in the world-line formalism.
Our starting point is the second order description for the
fermionic one-loop effective action.
The expression for the effective action in the second order
formulation allows us to justify the expressions for the effective
action in the world-line formalism which we propose later on.
Whereas in
the examples of ref.~\cite{mnss:hd-thep-94-51} only even numbers of outer
pseudo-scalars were allowed%
\footnote{We thank D.~Geiser-Gagn\'e for pointing this out to us.}
here we do not have that restriction.
Among the examples we present here, we also compute the chiral anomaly.

Let us begin with the second order description for the
one-loop effective action of a fermion in an arbitrary background. The
main difference between the usual (first order) description of a Dirac
particle and the second oder description is in the form of the
propagator. The first order propagator for a Dirac particle in
Euclidean space is
$i/({\not} p+m)$. In the second order description, one studies
basically the square of the Dirac equation. Therefore, the propagator
is now of Klein-Gordon type $1/(p^2+m^2)$.
The space-time propagator generated in the world-line formalism (and
string theory) is usually of the second type, independent of the nature
of the propagating field. For this reason, the second order
formalism~\cite{hos:jmp26,bd:npb379,mor:plb351}
is an important link in establishing the connection between
expressions in the world-line formalism and ordinary Feynman diagram
results~\cite{bd:npb379,mnss:hd-thep-94-51}.

In Euclidean space, the Lagrangian for a Dirac particle is given by
\begin{equation}
    {\cal L} = \bar\psi {\cal O} \psi,
  \label{dirac-e}
\end{equation}
where
\begin{equation}
    {\cal O} = (\partial_\mu + i gV_\mu + i g_5 \gamma_5 A_\mu) \gamma^\mu
                - i m - i \lambda\phi + \gamma_5 \lambda'\phi',
  \label{dirac-o}
\end{equation}
with the convention that~%
$\{\gamma_\mu,\gamma_\nu\} = -2\delta_{\mu\nu}$
and~$\gamma_\mu^\dagger = -\gamma_\mu$.

The corresponding one-loop effective action is
\begin{equation}
  \Gamma = \Tr \log {\cal O}
         = \frac{1}{2}\left\{
            \Tr(\log{\cal O} + \log{\cal O}')
            + \Tr(\log{\cal O} - \log{\cal O}')
           \right\}.
  \label{dirac-ea}
\end{equation}
To arrive at a second order formulation,
one can pick an arbitrary operator~${\cal O}'$ with the
restriction that the free part in~${\cal O}{\cal O}'$ is
quadratic. However, a good choice of~${\cal O}'$ is important for an
efficient perturbation theory derived from~(\ref{dirac-ea}). In
particular, it is convenient if we can choose~${\cal O}'$ such that
the second term vanishes. It is also convenient if it can be arranged
such that in~${\cal O}{\cal O}'$ no covariant derivative acts to the right,
except for those in the kinetic term.

For a general~${\cal O}$ as given in~(\ref{dirac-o}), there is no
choice for the operator~${\cal O}'$ which satisfies both criteria.
For the usual
choice~${\cal O}' = \gamma_5{\cal O}\gamma_5$, the second term
vanishes since~$\det{\cal O} = \det{\cal O}'$. However,  in the
presence of axial and pseudo-scalar couplings, the second
criterion mentioned above cannot be satisfied at the same time.
This makes the
translation of the expressions in the second order formalism into
expressions in the world-line formalism difficult.
Instead we choose
\begin{equation}
  {\cal O}' \equiv
  {\cal O}^\dagger =
    (\partial_\mu + i gV_\mu - ig_5 \gamma_5 A_\mu) \gamma^\mu
                      +i m + i \lambda\phi + \gamma_5 \lambda'\phi'.
  \label{o-prime}
\end{equation}
As we will see later on, this choice makes it straightforward to
translate~(\ref{dirac-ea}) into a world-line expression.

Rewriting the first term in~(\ref{dirac-ea}) as~$\log {\cal OO}^\dagger$, we
see that the
operator which generates the real part of the effective
action in the second order formalism is
\begin{equation}
  \begin{array}[b]{RL}
    {\cal OO}^\dagger =&
        {-} D_\mu D^\mu + g \sigma^{\mu\nu} V_{\mu\nu}
          + g_5 \gamma_5 \sigma^{\mu\nu} A_{\mu\nu}
          + i\lambda\gamma^\mu\partial_\mu\phi
          - \gamma_5 \lambda'\gamma^\mu\partial_\mu\phi'\\
        & + 2 (i m + i \lambda\phi - \gamma_5\lambda'\phi')
              i g_5 \gamma_5 A_\mu \gamma^\mu
          + m^2 + 2 m \lambda \phi + \lambda^2 \phi^2 + \lambda'^2 \phi'^2,
  \end{array}
  \label{oo-prime}
\end{equation}
where $D_\mu = \partial_\mu + i gV_\mu + i g_5 \gamma_5 A_\mu$.
We use~$V_{\mu\nu}$ and~$A_{\mu\nu}$ to denote the field-strength tensors.

The term which is a bit more difficult to handle is the second term
in~(\ref{dirac-ea}), corresponding to the imaginary part of the
effective action~$\Gamma$.
The imaginary part of~$\Gamma$ is generated by processes involving at
least one axial vector or one pseudo-scalar. This follows immediately
from the fact that for~${\cal O}'=\gamma_5 {\cal O} \gamma_5$ the
second term vanishes while the first term gives the same result as
for~${\cal O}'={\cal O}$.
If we view the effective action primarily as the
generating functional for one-particle irreducible Green's functions
the next step is clear: Instead of looking at the original term we
look at the term after one functional differentiation.
Using
\begin{equation}
  \frac{\delta}{\delta U} \Tr(\log{\cal O} - \log{\cal O}') =
    \Tr\left(
      \frac{\delta{\cal O}}{\delta U} \frac{1}{{\cal O}}
      - \frac{\delta{\cal O}'}{\delta U} \frac{1}{{\cal O}'}
    \right)
  \label{diff}
\end{equation}
and the cyclicity of the trace we can rewrite the derivative of the second term
of~(\ref{dirac-ea}) as
\begin{equation}
  \Tr\left(\left\{
      \frac{\delta{\cal O}}{\delta U} {\cal O}'
      - {\cal O} \frac{\delta{\cal O}'}{\delta U}
    \right\}
    \frac{1}{{\cal OO}'}
  \right)
  \label{diff-2o}
\end{equation}
where $U$ is either $A_\mu$ or $\phi'$.
This way, the expression we have to study is of the form of some
operator times a second order propagator. From
equations~(\ref{oo-prime},~\ref{diff-2o}) we can construct Feynman rules
for their perturbative evaluation
(tables~\ref{tab:2o-feynman},~\ref{tab:2o-feynman-c}).%
\begin{table}[tp]
  \newcolumntype{F}{l>{$}l<{$}|}
  \begin{center}
    \leavevmode
    \begin{tabular}{|*{2}{F}}
      \hline
      \vffb{V_\mu}& g(p_\mu+2q_\mu - i\sigma_{\mu\nu} p_\nu)&
      \vffb{A_\mu}& g_5\gamma_5(p_\mu+2q_\mu - i\sigma_{\mu\nu} p_\nu
                                -2m\gamma_\mu)\\
      \hline
      \vffb{\phi}& -\lambda (p_\nu \gamma^\nu + 2m)&
      \vffb{\phi'}& -i\lambda' \gamma_5 p_\nu \gamma^\nu\\
      \hline
      \vffbb{V_\mu}{V_\nu}& - 2 g^2 \delta_{\mu\nu}&
      \vffbb{A_\mu}{A_\nu}& - 2 g_5^2 \delta_{\mu\nu}\\
      \hline
      \vffbb{A_\mu}{\phi}& 2\lambda g_5 \gamma_5 \gamma_\mu&
      \vffbb{A_\mu}{\phi'}& 2i\lambda' g_5 \gamma_\mu\\
      \hline
      \vffbb{\phi}{\phi}& -2\lambda^2&
      \vffbb{\phi'}{\phi'}& -2\lambda'^2\\
      \hline
      \multicolumn{2}{|r}{\vffbb{V_\mu}{A_\nu}}&
      \multicolumn{2}{l|}{$- 2 g g_5 \gamma_5 \delta_{\mu\nu}$}\\
      \hline
    \end{tabular}
    \caption{Feynman rules for the modified second order
      formalism~(\protect\ref{oo-prime}).
      We use~$p$~($=i\partial$) to denote the momentum of the incoming boson,
      and~$q$ for the momentum of the incoming fermion.
      A global factor of~$1/2$ and a negative sign for the fermion
      loop are required.}
    \label{tab:2o-feynman}
  \end{center}
\end{table}%
\begin{table}[tp]
  \newcolumntype{F}{l>{$}l<{$}|}
  \begin{center}
    \leavevmode
    \begin{tabular}{|*{2}{F}}
      \hline
      \vffb{A_\mu}& - g_5\gamma_5 (i\sigma_{\mu\nu} (p_\nu + 2q_\nu) - p_\mu)&
      \vffb{\phi'}& -i\lambda' \gamma_5 (\gamma_\nu (p_\nu + 2 q_\nu) - 2m)\\
      \hline
      \vffbb{A_\mu}{A_\nu}& 2 i g_5^2 \sigma_{\mu\nu}&
      \vffbb{\phi'}{V_\nu}& 2i\lambda' g \gamma_5 \gamma_\nu\\
      \hline
      \vffbb{A_\mu}{V_\nu}& 2 igg_5 \gamma_5 \sigma_{\mu\nu}&
      \vffbb{\phi'}{\phi}& -2 \lambda \lambda' \gamma_5\\
      \hline
    \end{tabular}
    \caption{Feynman rules for the extra terms
      (braces in eq.~(\protect\ref{diff-2o})).
      Besides the ordinary second order calculation, one also has to
      compute diagrams where one vertex is taken from this table and
      all others are ordinary vertices from
      table~\protect\ref{tab:2o-feynman}.}
    \label{tab:2o-feynman-c}
  \end{center}
\end{table}

The perturbative evaluation of~(\ref{oo-prime}) using the rules from
table~\ref{tab:2o-feynman} is the standard procedure. To evaluate
the extra terms generated by~(\ref{diff-2o}), one has to take one
vertex from table~\ref{tab:2o-feynman-c} (corresponding to the
term in braces) and all others from table~\ref{tab:2o-feynman}. All
propagators are of Klein-Gordon type. Using this set of rules it is
straightforward to compute the one-particle irreducible
one-loop Green's functions.

\begin{figure}[tp]
  \begin{center}
    \leavevmode
    \begin{tabular}{c@{\hspace*{3cm}}c}
    \epsffile{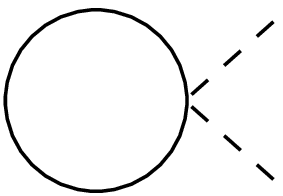} & \epsffile{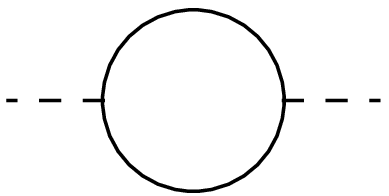}\\
     (a)                 & (b)
    \end{tabular}
  \end{center}
  \caption{Two point function in the second order formalism. One
    always has two terms: A diagram with a contact interaction and a
    diagram with two three-point interactions.}
  \label{fig:2pt}
\end{figure}
Now let us look at some examples.
For the calculations we present here, we use dimensional
regularisation. We verified, however, that the same results are
obtained using %
Pauli-Villars regularisation%
. %
Since in this paper, we only deal with one-loop
processes, we use a na{\"\i}ve scheme for the treatment
of~$\gamma_5$ in dimensional regularisation~\cite{cfh:npb159}.

To warm up, let us compute the axial-axial two-point function. In
principle, we have to compute four terms: The regular term
(eq.~(\ref{oo-prime})) and the extra term (eq.~(\ref{diff-2o})) for
diagrams~(a) and~(b) of figure~\ref{fig:2pt}. As it turns out, only
the regular terms contribute while the extra terms vanish. The result
of this calculation is for arbitrary, even dimension
\begin{equation}
  \begin{array}[b]{RR}
    \langle A_\mu A_\nu \rangle_{(\mathrm{a})} =&
      \multicolumn{1}{L}{
        n_F g_5^2 (4\pi)^{-d/2} (m^2)^{d/2-1} \Gamma(1-d/2) \delta_{\mu\nu}
      }\\
    \langle A_\mu A_\nu \rangle_{(\mathrm{b})} =&
      \frac{n_F}{2}g_5^2 (4\pi)^{-d/2} \int_0^1 \rmd x \left\{
        -\Gamma(1-d/2) 2 \delta_{\mu\nu} (x(1-x) p^2 + m^2)^{d/2-1}
      \right. \\
        &+\Gamma(2-d/2) (-\delta_{\mu\nu} (p^2 + 4m^2)
                         + 4x(1-x) p_\mu p_\nu) \times
      \\
        &\times\left.(x(1-x)p^2 + m^2)^{d/2-2} \right\},
  \end{array}
  \label{AA}
\end{equation}
where $n_F=\tr(1)$ is the number of fermionic degrees of freedom.
This is related to the standard first order result
\begin{equation}
  \begin{array}[b]{RR}
    \langle A_\mu A_\nu \rangle =&
      {-}2 n_F g_5^2 (4\pi)^{-d/2} \int_0^1 \rmd x\, \Gamma(2-d/2)
        (x(1-x)p^2 +m^2)^{d/2-2} \times\quad\\
      &\times\left\{g_{\mu\nu}(x(1-x)p^2 +m^2) - x(1-x) p_\mu p_\nu\right\}
  \end{array}
  \label{AA-ft}
\end{equation}
by integration by parts and by $\Gamma$-function identities.

A more interesting example is given by the axial-vector and vector
pseudo-scalar two-point functions in two dimensions. With those two
processes we can study the axial current Ward identity and the chiral
anomaly in two dimensions.
It turns out that for this case, the contribution from the
regular second order expression vanishes.
Instead, the process is described using the extra Feynman rules
(table~\ref{tab:2o-feynman-c}). Here, we chose the vertex to be a
special vertex for the coupling to the axial current. The other vertex
for the vector current is an ordinary vertex. Besides this,
the calculation uses standard Feynman techniques.
\begin{equation}
  \begin{array}[b]{RR}
    \langle A_\mu V_\nu \rangle_{(\mathrm{a})} =&
      \multicolumn{1}{L}{
        {-}2 igg_5 (4\pi)^{-d/2} (m^2)^{d/2-1} \Gamma(1-d/2) \epsilon_{\mu\nu}
      }\\
    \langle A_\mu V_\nu \rangle_{(\mathrm{b})} =&
      igg_5 (4\pi)^{-d/2} \int_0^1 \rmd x \left\{
         \Gamma(1-d/2) 2 \epsilon_{\mu\nu} (x(1-x) p^2 + m^2)^{d/2-1}
      \right. \\
        &+\Gamma(2-d/2) (\epsilon_{\mu\nu} p^2
                         -4 x(1-x) \epsilon_{\mu\rho} p_\rho p_\nu) \times
      \\
        &\times\left.(x(1-x)p^2 + m^2)^{d/2-2} \right\}.
  \end{array}
  \label{AV}
\end{equation}
and
\begin{equation}
  \langle P V_\nu\rangle =
    -2 m g\lambda' \epsilon_{\mu\nu} p_\mu \Gamma(2-d/2)
      \int_0^1 (x(1-x) p^2 + m^2)^{d/2-2}.
  \label{PV}
\end{equation}
Again, these expressions are equivalent to the results of an ordinary,
first order calculation.
If we check the axial current Ward identity
\begin{equation}
  \frac{i}{g_5}p_\mu \langle A_\mu V_\nu \rangle =
                         \frac{2m}{\lambda'}\langle P V_\nu\rangle
  \label{ward}
\end{equation}
we find, as expected,  on the right hand side an extra, anomalous term
\begin{equation}
  4g (4\pi)^{-d/2} \int_0^1 \rmd x\,
    \Gamma(2-d/2) \epsilon_{\mu\nu} p_\mu (x(1-x) p^2 + m^2)^{d/2-1}.
  \label{anomaly}
\end{equation}
A first order field-theory calculation verifies that this is indeed the
axial anomaly in two dimensions.

Our goal now is to translate these results into a world-line
formulation. Since we study the couplings to an internal fermion we
start from the usual description of a spinning particle by a
supersymmetric world-line
Lagrangian~\cite{bdzdh:pl64b,bdh:npb118,bm:ap104} using a curved
superspace description. The particle is described by a superfield
$
  X^\mu{}(\tau{},\theta{})
    = x^\mu{}(\tau{}) + \theta{}\sqrt{e}\,\psi{}^\mu{}(\tau{}),
$
where $x$ is a normal commuting number, and $\theta{}$ is a Grassmann
variable. To keep the manifest reparametrisation invariance of the
super world-line, we introduce the super-einbein
$
  \Lambda{} = e + \theta{}\sqrt{e}\,\chi{}.
$
Furthermore, to couple the spinning particle to scalar and
pseudo-scalar fields, we need two auxiliary fields
$
\bar X, X' = \sqrt{e}\, \psi_{5,6} + \theta x_{5,6}
$%
{}~\cite{mnss:hd-thep-94-51}.
Since we deal with a curved super-space on the world-line, we also have to
distinguish the two derivatives
$
  D_\theta{} = \Lambda{}^{-1/2}
                 \left(\frac{\partial}{\partial\theta{}}
                       -
                       \theta{}\frac{\partial}{\partial\tau{}}\right)
$ and
$
  D_\tau{} = \Lambda{}^{-1}\frac{\partial}{\partial\tau{}}.
$
Translating the second order action~(\ref{oo-prime}) into world-line
form we suggest the
world-line action
\begin{equation}
  \begin{array}[b]{RR}
    S=&\int \rmd\tau\rmd\theta\,\Lambda^{1/2} \bigg(
       \frac{1}{2}D_\tau X \cdot D_\theta X +
       \frac{1}{2} \Lambda^{-1} \bar X \cdot D_\theta \bar X +
       \frac{1}{2} \Lambda^{-1} X' \cdot D_\theta X +\qquad\\
      &{}- g_5 \bar X X' D_\theta X_\mu A_\mu +
       \Lambda^{-1/2} i\lambda' X' \Phi' + \quad\\
      &{}+ig D_\theta X_\mu V_\mu +
       \Lambda^{-1/2} i\lambda \bar X \Phi \bigg).
  \end{array}
  \label{action-s}
\end{equation}

Before we can do any calculations using this action, we have to fix
the reparametrisation invariance by fixing~$\Lambda$.
For calculations on the circle, a valid and convenient gauge is~$e=2$,
$\chi=0$.
In this gauge, the world-line action in component notation is
\begin{equation}
  \begin{array}[b]{RL}
  S = \int \rmd\tau{} \bigg(&
                  \frac{\dot x^2}{4}
                  +  \frac{x_5^2}{4}
                  +  \frac{x_6^2}{4}
                  +  \frac{1}{2} \psi{}\dot\psi{}
                  +  \frac{1}{2} \psi_5\dot\psi_5
                  +  \frac{1}{2} \psi_6\dot\psi_6
                 \\[0.5ex]
                 &+ 2 i\lambda( x_5 \Phi
                              - 2 \psi_5 \psi \cdot \partial \Phi)
                  + 2 i\lambda'( x_6 \Phi'
                              - 2 \psi_6 \psi \cdot \partial \Phi')\\[0.3ex]
                 &+ g_5 2 \Big[\psi_5\psi_6(\dot x_\mu A_\mu
                                    + 2 \psi_\mu \psi_\nu \partial_\nu A_\mu)
                                  +(\psi_5 x_6 - \psi_6 x_5) \psi_\mu A_\mu
                             \Big]\\[0.3ex]
                 &- ig (\dot x_\mu V_\mu
                        + 2 \psi_\mu \psi_\nu \partial_\nu V_\mu)
                  \bigg).
  \end{array}
\label{wla}
\end{equation}
A mass term for the fermion is generated by shifting
$\phi\to\phi+m/\lambda$~\cite{mnss:hd-thep-94-51}.

Having this world-line action is not all.
For example, one can check immediately that it is impossible to
generate diagrams with an odd number of pseudo-scalars.
This corresponds to the situation in the second order formalism, so
we also need to describe the terms corresponding to~(\ref{diff}).
They can all be generated from
\begin{equation}
  \Gamma'_U = \frac{\delta}{\delta U}i\mathop{\Im\textit{m}} \Gamma
            = -\frac{n_F}{2} \int_0^\infty \frac{\rmd T}{T}
                 \int {\cal D}X {\cal D}\bar X {\cal D}X'
                   (-1)^F \Omega_U e^{-S}
  \label{diff-wl}
\end{equation}
where we have~$U=A_\mu$ or~$\phi'$.
We can translate the terms in table~\ref{tab:2o-feynman-c} in a
suggestive manner.
The inserted operators are then
\begin{equation}
  \begin{array}[b]{LLL}
    \Omega_A
      &\multicolumn{2}{L}{=-g_5 \int_0^T \rmd\tau\,
           \biggl(\psi_\mu \psi_\nu \dot x_\nu
                  - \frac{i}{2} p_\mu\biggr)\psi_5\psi_6}\\
      &= -g_5 \int_0^T \rmd\tau\rmd\theta\,\theta&
      \biggl\{
      \biggl(
        D_\tau X \cdot D_\theta X +
        \Lambda^{-1} \bar X \cdot D_\theta \bar X +
        \Lambda^{-1} X' \cdot D_\theta X'
      \biggr)
        D_\theta X_\mu\\
      &&\qquad- \Lambda^{-1} i p_\mu\biggr\} \bar X X'
    \\
    \Omega_{\phi'}
      &\multicolumn{2}{L}{= -i\lambda' \int_0^T \rmd\tau\,
         \biggl(\frac{1}{2}\psi_\nu\dot x_\nu
          + \frac{1}{2}\psi_5 x_5\biggr)\psi_6
        }\\
      &= -i\lambda' \int_0^T \rmd\tau\rmd\theta\,\theta&
      \biggl(
        D_\tau X \cdot D_\theta X +
        \Lambda^{-1} \bar X \cdot D_\theta \bar X +
        \Lambda^{-1} X' \cdot D_\theta'
      \biggr)
      X'
  \end{array}
  \label{diff-wl-insertions}
\end{equation}
The quartic vertices in table~\ref{tab:2o-feynman-c} are generated by
$\delta$-functions which appear in some of the world-line Green's
functions. The explicit~$p_\mu$ in the terms is really a derivative
acting on~$U$. %

The operator~$(-1)^F$ anti-commutes with all fermionic fields and this
way implements the usual~$\gamma_5$ in the Dirac algebra.
The presence of~$(-1)^F$ changes the boundary conditions for
world-line fermions from anti-periodic to
periodic and is known to play a r\^ole in the computation of
anomalies~\cite{a-g:cmp90,aw:npb234}.
This and the introduction of the auxiliary fields~$X'$ and~$\bar X$
are the only new ingredients in the action.

All these terms are of a surprisingly simple form, which is almost
a product of the
kinetic part of the Lagrangian in~(\ref{wla}) with the interaction
part of the field under consideration, all projected to the lower
component of the superfield.
This is not manifestly supersymmetric.

Before we can perform any calculations in this world-line formalism,
we have to expand the action in components and perform the integral over
the auxiliary fields. This step is necessary to resolve some
ambiguities in the evaluation of products of Green's functions on the
world-line (Similar ambiguities appeared in the world-line
calculations of~\cite{bpsn:npb446}).
These ambiguities arise since~$\ddot G_B$
and~$\langle x_5 x_5\rangle$ contain a
$\delta$-function while~$\dot G_B$ and~$G_F$ contain a
step-function. After the removal of the auxiliary fields from the
world-line action, we can always use~$G_F^2=1$ in all calculations,
even for identical arguments (relevant if multiplied by a
$\delta$-function). An example of this ambiguity occurs in the
calculation of the axial current two-point function. We see that both
the~$\psi_5\psi_6\dot x_\mu A_\mu$ term and
the~$(\psi_5 x_6 - \psi_6 x_5)\psi_\mu A_\mu$ term in~(\ref{wla})
generate~$G_F^2(u) \delta(u)$.
However, to reproduce the second order calculation, only the first
term is allowed to contribute while the contribution from the second term
has to vanish.

Integrating out the auxiliary fields removes
the second term in the axial coupling in~(\ref{wla}) and introduces
quartic couplings
instead. The resulting world-line action bears a much closer relation
to the second order expression for the effective
action~(\ref{oo-prime}). Some of the subtleties mentioned here will be
further discussed in a forthcoming publication~\cite{mnss:wip}.

Now let us look at our examples again. The evaluation of the axial-vector
axial-vector two-point function is fairly simple. The calculation
follows the usual
world-line rules~\cite{str:npb385,ss:plb318,mnss:hd-thep-94-51}.
For the axial vector two-point
function we find
\begin{equation}
  \begin{array}[b]{RL}
    \langle A_\mu A_\nu \rangle = &
      \frac{n_F}{2}g_5^2 \int_0^\infty \frac{\rmd T}{T} (4\pi T)^{-d/2} T^2
        \int_0^1 \rmd u \Big\{-\delta_{\mu\nu}(p^2 - \frac{1}{T} \ddot G_B
                                          +4m^2)\\
    &\multicolumn{1}{R}{
                 \qquad - p_\mu p_\nu( (\dot G_B)^2 - 1)\Big\}
        \exp(-T(m^2 + p^2 G_B))}
  \end{array}
  \label{AA-wl}
\end{equation}
where~$G_B$ is the bosonic Green's function on the circle.
Replacing~$G_B = u(1-u)$, $\dot G_B=1-2u$, and $\ddot G_B=2\delta(u)-2$
it is easy
to check that this is indeed the result from eq.~(\ref{AA}).
In the Green's function, the~$\delta$-function is understood to be on
the circle.
As in the second order calculation, the extra terms do not play a r\^ole
in this process.

More interesting is the computation of the vector-axial vector and
pseudo-scalar vector two-point functions. We can see immediately that the
direct term generated by~(\ref{AA-wl}) vanishes.
So we have to look at the terms generated by~(\ref{diff-wl}).
Here, the~$(-1)^F$ appears, which corresponds to the presence
of a~$\gamma_5$ in the field-theory calculation.
This means that the boundary conditions for the fermions get changed
and both fermions and bosons have periodic boundary conditions.
In this case there
are zero modes for the fermions which we have to take into
account. Furthermore, the fermionic Green's function is changed and we
have
\begin{equation}
  G_F^{(\rm periodic)} = \dot G_B,
  \label{GF}
\end{equation}
{\em i.~e.}, the world-line supersymmetry is no longer broken by the
boundary conditions. We get now
\begin{equation}
  \begin{array}[b]{RL}
    \langle A_\mu V_\nu \rangle = &
      {-}igg_5\int_0^\infty \frac{\rmd T}{T} (4\pi T)^{-d/2} e^{-Tm^2}\times\\
       &\times T^2\int_0^1 \rmd u\,\Bigl[
         \langle\psi_\mu\psi_\rho\rangle''
         \langle \dot x_{\rho,1} \dot x_{\nu,2}
                                    e^{-ipx_1} e^{ipx_2}\rangle\\
       &\qquad+\langle\psi_\nu\psi_\rho\rangle'' p_\mu p_\rho
                                    \langle e^{-ipx_1} e^{ipx_2}\rangle\\
       &\qquad+2ip_\rho\langle\psi_{\mu,1}\psi_{\rho,1}
                         \psi_{\nu,2}\psi_{\delta,2}\rangle''
           \langle\dot x_{\delta,2} e^{-ipx_1} e^{ipx_2}\rangle
         \Bigr].
  \end{array}
  \label{AV-wl-u}
\end{equation}
Here the extra subscripts 1, 2 denote the vertex insertion which the
fields belong to.
Evaluating this term we notice that we have to include one zero mode
for each fermion present. For~$\psi_5$ and~$\psi_6$ this is already
done. For the~$\psi_\mu$, this is indicated by the modified
contraction~$\langle\rangle''$. In~$n$ dimensions this automatically produces a
factor~$\prod_{i=1}^n \psi_{\mu_i}$ which is nothing but the
$n$-dimensional $\epsilon$-tensor. The result is then
\begin{equation}
  \begin{array}[b]{RL}
    \langle A_\mu V_\nu\rangle =
      {-}igg_5&\int_0^\infty\frac{\rmd T}{T} (4\pi T)^{-d/2} e^{-Tm^2}
                                                               T^2 \times
        \\
      &\times\int_0^1 \rmd u\, \Bigl[
          \epsilon_{\mu\nu}\ddot G_B
          - \epsilon_{\mu\alpha} p_\alpha p_\nu \dot G_B^2
          + \epsilon_{\nu\beta} p_\mu p_\beta
        \Bigr] e^{-TG_B p^2},
  \end{array}
  \label{AV-wl}
\end{equation}
where we used the identity~%
$
\epsilon_{\mu\nu}p^2 = \epsilon_{\mu\alpha} p_\alpha p_\nu
                       - \epsilon_{\nu\beta} p_\mu p_\beta
$.
We use this identity again and substitute the world-line Green's
functions to recover the second order result~(\ref{AV}).
For the pseudo-scalar vector coupling we find
\begin{equation}
  \begin{array}[b]{RL}
    \langle P V_\nu \rangle
      &=2\lambda' g m p_\rho
        \int_0^\infty\frac{\rmd T}{T} (4\pi T)^{-d/2} e^{-Tm^2}
        T^2\int_0^1 \rmd u\, \langle\psi_\nu\psi_\rho\rangle''
        e^{-TG_B p^2}\\
      &=-2\lambda' g m \epsilon_{\rho\nu} p_\rho
        \Gamma(2-d/2) \int_0^1 \rmd u\,
        (m^2 + u(1-u) p^2)^{d/2-2}.
  \end{array}
  \label{PV-wl}
\end{equation}
Both expressions agree with the second order results obtained
above. This implies that also in the world-line formalism we were able
to compute both the axial current Ward identity and the axial anomaly.
It is possible, even though slightly more cumbersome, to do the same
calculation using a Pauli-Villars regulator instead of dimensional
regularisation. The final result is not affected by this choice.

In four dimensions, we can do the same for the triangle graphs with one
or three axial currents~\cite{mnss:wip}. Again, the $\epsilon$-tensors
are produced by the zero-modes of the fermion fields.

We have seen how useful the second order formalism is
as a tool for
constructing of world-line actions for a spinning particle in a loop. By
constructing an appropriate second order representation for a
theory with Dirac fermions we were able to find a world-line
representation which reproduces exactly the second order expressions
we started with.
This indicates strongly that the world-line formalism
suggested here indeed reproduces the usual field theory results.
An important new ingredient are auxiliary fields~$X'$ and~$\bar X$,
and the inclusion
of~$(-1)^F$ in the representation of~$\gamma_5$. This allowed us to
treat processes with an odd number of vertices with
$\gamma_5$-couplings, especially those connected to the chiral anomaly.
A computerized higher order effective action calculation to verify the
agreement between the standard approach and the world-line formalism
is in preparation~\cite{mnss:wip}.
Furthermore it is interesting to see if this kind of world-line
formulation of axial couplings allows for the generalization to
multi-loop processes and to processes with open fermion lines.

In a recent preprint~\cite{dg:hep-th-9508131}, which we received while
finishing this letter, D'Hoker and Gagn\'e also introduce the
operator~$(-1)^F$ for the calculation of the imaginary part of the
effective action as a mean to generate the $\epsilon$-tensor from the
fermionic zero modes. They do not include axial vector couplings, though.
Furthermore, they
derive an elegant integral representation for the imaginary part of
the effective action. This representation has the advantage of
providing a closed
form for the imaginary part which is still missing in our approach.
The resulting perturbation theory  introduces an additional Feynman
parameter-like integration.
It seems to us that the construction of the world-line representation as we
present it here, including the superfield formulation, can also be
done starting from their integral representation.

An alternative construction of the Yukawa coupling of a spinning
particle and a bosonic representation of~$\gamma_5$ are given
in~\cite{hol:hep-th-9508136}.

\noindent
{\bf Acknowledgments:} We acknowledge Jan Willem van Holten for
useful discussions, and Jan Eeg for
drawing our attention to the relevance of effective actions involving
axial currents and for early collaboration.
One of us
(C. S.) would like to thank the Aspen Center for Physics for
hospitality.

\newcommand{\noopsort}[1]{} \newcommand{\switchargs}[2]{#2#1}

\end{document}